\documentstyle[aps,prl,multicol,epsf,graphics]{revtex}

\begin{document}

\title{Griffiths-McCoy singularities in random quantum spin chains:\\
Exact results through renormalization}

\author{Ferenc Igl\'oi$^{1,2,3}$, R\'obert Juh\'asz$^{2,1}$ and P\'eter Lajk\'o$^{2}$}

\address{
$^1$ Research Institute for Solid State Physics and Optics, 
H-1525 Budapest, P.O.Box 49, Hungary\\
$^2$ Institute for Theoretical Physics,
Szeged University, H-6720 Szeged, Hungary\\
$^3$Laboratoire de Physique des Materiaux, Universit\'e Henri Poincar\'e (Nancy 1),
F-54506 Vand\oe uvre les Nancy, France
}

\date{April 28, 2000}

\maketitle

\begin{abstract}
The Ma-Dasgupta-Hu renormalization group (RG) scheme is used to study singular quantities
in the Griffiths phase of random quantum spin chains. For the random transverse-field
Ising spin chain we have extended Fisher's analytical solution to the off-critical region
and calculated the dynamical exponent exactly. Concerning other random chains we argue by
scaling considerations that the RG method generally becomes asymptotically exact for
large times, both at the critical point and in the whole Griffiths phase. This statement
is checked via numerical calculations on the random Heisenberg and quantum Potts models
by the density matrix renormalization group method.
\end{abstract}


\newcommand{\bc}{\begin{center}}
\newcommand{\ec}{\end{center}}
\newcommand{\be}{\begin{equation}}
\newcommand{\ee}{\end{equation}}
\newcommand{\beqn}{\begin{eqnarray}}
\newcommand{\eeqn}{\end{eqnarray}}

\begin{multicols}{2}
\narrowtext
In a random quantum system at zero temperature several physical quantities are singular not only at the
critical point, but in a whole region, as well, which extends on both sides of the transition point\cite{bhatt}.
In this Griffiths phase\cite{griffiths} a random quantum system is non-critical in the space direction
(spatial correlations decay exponentially), whereas it is critical in the time direction and the
corresponding behavior due to Griffiths-McCoy singularities\cite{griffiths,mccoy} is controlled by a line of
semi-critical fixed points characterized by the dynamical exponent $z(\delta)$, which depends on the
value of the quantum control-parameter, $\delta$. For example average autocorrelations decay in (imaginary) time, $\tau$,
as $\sim \tau^{-z(\delta)}$; for a small magnetic field, $H \to 0$, the magnetization behaves as $\sim H^{1/z(\delta)}$;
the low temperature susceptibility and specific heat are singular as $\sim T^{1/z(\delta)-1}$ and $\sim T^{1/z(\delta)}$,
respectively. Let us mention that
Griffiths-McCoy singularities are relevant in experiments on quantum
spin glasses\cite{exp} and provides a theoretical explanation about the non-Fermi liquid behavior in
U and Ce intermetallics\cite{neto}.

Among the theoretical methods developed to study random quantum systems the renormalization group (RG) scheme
introduced by Ma, Dasgupta and Hu\cite{mdh} plays a special r\^ole. For a class of systems, the critical behavior
of those is controlled by an infinite-randomness fixed point\cite{2d} (IRFP), the RG method becomes
asymptotically exact during iteration. For some one-dimensional models, random transverse-field
Ising model\cite{fisher} (RTIM) and the random XXZ-model\cite{fisherxx}, Fisher has obtained analytical solution
of the RG equations and in this way many new exact results and new physical insight about the critical
behavior of these models have been gained. Subsequent analytical\cite{bigpaper} and
numerical\cite{num,bigpaper} investigations of the models are in agreement with Fisher's results. The RG-scheme
has been numerically implemented in higher dimensions\cite{2d,yucheng}, as well, to study the critical behavior of the RTIM and
reasonable agreement with the results of quantum Monte-Carlo simulations\cite{mc} has been found.
Considering the Griffiths-phase of random quantum spin chains here the RG-scheme has been rarely used\cite{yucheng}, mainly
due to the general belief that the method looses its asymptotically exact properties by leaving the vicinity of
the scale invariant critical point. 

Our aim in the present Letter is to clarify the applicability of the Ma-Dasgupta-Hu RG-method
in the Griffiths phase of random quantum spin chains. We start with the RTIM, extend Fisher's calculation\cite{fisher}
to the Griffiths phase of the model and present the analytical solution of the RG-equations. Then, for general models,
we analyze by scaling considerations the structure of the RG equations around the line of semi-critical fixed points and arrive to
the conclusion that the RG method becomes asymptotically exact in the whole Griffiths region.
This statement is then checked numerically
on the random Heisenberg chain and the random quantum Potts model (RQPM) using the density matrix renormalization group (DMRG)
method.

We start with the 1d RTIM which is defined by the Hamiltonian:
\be
H_I=-\sum_i J_i \sigma_i^x\sigma_{i+1}^x- \sum_i h_i \sigma_i^z\;,
\label{hamilton}
\ee
where the $\sigma_i^{x,z}$ are Pauli matrices at site $i$ and $J_i$ and $h_i$ are the couplings
and the transverse fields, respectively, which are independent random variables. The quantum control-parameter
of the model is defined as
\be
\delta={[\ln h]_{\rm av} - [\ln J]_{\rm av} \over {\rm var}[\ln h]+{\rm var}[\ln J]}\;,
\label{delta}
\ee
where ${\rm var}[x]$ is the variance of $x$ and we use $[\dots]_{\rm av}$ to denote averaging over
quenched disorder. For $\delta>0$ ($\delta<0$)
the system is in the paramagnetic (ferromagnetic) phase, so that the random quantum critical
point is at $\delta=0$.

In the Ma-Dasgupta-Hu RG-method the strongest term in the
Hamiltonian, coupling or field, of strength $\Omega$ are successively decimated out and the
neighboring fields or couplings
are replaced by weaker ones, which are generated by a perturbation calculation. 
The basic RG-equations for coupling and field decimations are given by:
\be
\tilde{h}={h_i h_{i+1} \over J_i \kappa}\quad,\quad\tilde{J}={J_{i-1} J_i \over h_i \kappa}\;,
\label{htilde}
\ee
respectively, which are related through duality. Here, for the RTIM $\kappa=1$.
Under renormalization we follow the probability
distributions of the couplings, $R(J,\Omega)$, and that of the fields, $P(h,\Omega)$.
When the energy-scale is lowered as $\Omega \to \Omega - d \Omega$ the distribution of the couplings is
changed as:
\beqn
{{\rm d} R(J,\Omega) \over {\rm d} \Omega}&=&R(J,\Omega)[P(\Omega,\Omega)-R(\Omega,\Omega)] \nonumber \\
-P(&\Omega&,\Omega) \int_{J \kappa}^{\Omega} {\rm d} J' R(J',\Omega) R({J \Omega \kappa \over J'},\Omega)
{\Omega \kappa \over J'}\;,
\label{rgJ} 
\eeqn
where the first term in the r.h.s. is due to a balance between decimated couplings and
normalization, whereas the second term accounts for the generated new couplings.
A similar equation for the field distribution follows from Eq.(\ref{rgJ})
through duality
\beqn
{{\rm d} P(h,\Omega) \over {\rm d} \Omega}&=&P(h,\Omega)[R(\Omega,\Omega)-P(\Omega,\Omega)]  \nonumber \\
-R(&\Omega&,\Omega) \int_{h \kappa}^{\Omega} {\rm d} h' P(h',\Omega) P({h \Omega \kappa \over h'},\Omega)
{\Omega \kappa \over h'}\;,
\label{rgh} 
\eeqn
which amounts to interchange $J \leftrightarrow h$ and $R \leftrightarrow P$.

For the RTIM, i.e. for $\kappa=1$, we found one class of solutions of the RG equations
in the form:
\beqn
R(J,\Omega)&=&R(\Omega,\Omega) \left({\Omega / J}\right)^{1-R(\Omega,\Omega) \Omega} \quad \quad \nonumber \\
P(h,\Omega)&=&P(\Omega,\Omega) \left({\Omega / h}\right)^{1-P(\Omega,\Omega) \Omega}\ \quad ,
\label{solu}
\eeqn
where the distributions involve the parameters, $\tilde{R}(\Omega) \equiv R(\Omega,\Omega)$ and
$\tilde{P}(\Omega) \equiv P(\Omega,\Omega)$, which satisfy the relation
$(\tilde{P}-\tilde{R})\Omega={1 / z}=const$. Thus the solution is characterized by one parameter, $z=z(\delta)$,
which depends on the quantum control-parameter, $\delta$: at the critical point,
$1/z(0)=0$, whereas in the paramagnetic phase, $\delta>0$, $1/z(\delta)>0$ and monotonically increases
with $\delta$.

In terms of the variables, $y=\tilde{R}\Omega + 1/2z=\tilde{P}\Omega - 1/2z$ and $x=-\ln \Omega$
we obtain the differential equation:
\be
{{\rm d} y \over {\rm d} x} + y^2={1 \over 4 z^2}\;,
\label{yx}
\ee
which has the solution, $y=1/(x-x_0)$, $x_0={\rm const}$, {\it at the critical point} with $1/z=0$.
The distribution of $\rho=R \Omega$
in terms of the variable $\eta=-(\ln \Omega - \ln J)/\ln \Omega$ is given by $\rho(\eta)
{\rm d} \eta=\exp(-\eta) {\rm d} \eta$, which corresponds to the fixed-point solution
by Fisher\cite{fisher}. At this point we refer to Fisher's analysis\cite{fisher} and conclude that the functions
in Eqs(\ref{solu}) indeed represent the fixed-point distribution for all non-singular initial
distributions.

In the Griffiths phase, $\delta>0$, the solution of Eq.(\ref{yx}) in terms of the original energy-scale
variable, $\Omega$, is given by
\be
y={y_0{/ 2z}+{1 / 4z^2} \tanh\left[ \ln(\Omega_0/\Omega)/ 2z\right]
\over {1 / 2z}+y_0 \tanh\left[\ln(\Omega_0/\Omega)/ 2z\right]}\;,
\label{ysol}
\ee
where $y=y_0$ at a reference point, $\Omega=\Omega_0$. Approaching the line of
semi-critical fixed points, i.e. for $\Omega/\Omega_0 \to 0$, we have to leading order:
\be
\tilde{R} \Omega=\tilde{P} \Omega - {1 / z}={\tilde{R}(\Omega_0) /[
\tilde{P}(\Omega_0)} z] \left({\Omega / \Omega_0} \right)^{1/z} + ...
\label{appr}
\ee
thus $\tilde{P}$ and $\tilde{R}$ have different low energy asymptotics.

The {\it physical relevance} of $1/z$ can be obtained by studying the change
of number of spins, $n_{\Omega} \to n_{\Omega}-{\rm d}n_{\Omega}$ connected with a change in the energy scale
as $\Omega \to \Omega-{\rm d} \Omega$. This leads to the differential equation
${{\rm d} n_{\Omega} / {\rm d}\Omega}= n_{\Omega} \left[P(\Omega,\Omega)+R(\Omega,\Omega)
\right]$, the solution of which is given by:
\be
n_{\Omega}=\left\{ {\rm cosh}\left[ \ln(\Omega_0/\Omega)/ 2z\right]+
2z~y_0  {\rm sinh}\left[ \ln(\Omega_0/\Omega)/ 2z\right] \right\}^{-2}\;,
\ee
which along the line of semi-critical fixed points has the asymptotic behavior $n_{\Omega}= const~\Omega^{1/z}$,
$\Omega \to 0$. Since the typical distance between remaining spins is
$L_{\Omega} \sim {1 / n_{\Omega}} \sim \Omega^{-1/z}$,
we can identify $z$ as the {\it dynamical exponent}, which governs the relation between
time- and length-scales as $\tau \sim L^{z}$.

Next we show that $z$ is invariant along the RG trajectory and can be deduced from the
original distributions. For this we consider the
averages, $[J^{\mu}]_{\rm av}$ and $[h^{-\mu}]_{\rm av}$,
and using Eqs.(\ref{rgJ}) and (\ref{rgh}) we calculate the derivative:
\beqn
&.&{{\rm d} \over {\rm d} \Omega} \left[\left({J/h}\right)^{\mu}\right]_{\rm av}=
\left(1-\left[\left({J / h}\right)^{\mu}\right]_{\rm av}\right) \nonumber
\\  &\times&   \left(  P(\Omega,\Omega)\Omega^{-\mu}
[J^{\mu}]_{\rm av}  + R(\Omega,\Omega)\Omega^{\mu} [h^{-\mu}]_{\rm av} \right) 
\label{Jh}
\eeqn
which is vanishing for $\mu=\tilde{\mu}$, if $[(J/h)^{\tilde{\mu}}]_{\rm av}=1$. 
Consequently $\tilde{\mu}$ stays invariant along the RG trajectory until the fixed point, where using the distribution
in Eqs.(\ref{solu}) we obtain $\tilde{\mu}=1/z$.
Thus the dynamical exponent for the RTIM is given by the solution of the equation:
\be
\left[\left({J / h}\right)^{1/z}\right]_{\rm av}=1\;,
\label{zexact}
\ee
which is then exact, since the
RG-transformation becomes asymptotically exact as $\Omega \to 0$. This latter statement
follows from the fact that the ratio of decimated bonds, $\Delta n_J$, and decimated fields, $\Delta n_h$,
goes to zero as $\Delta n_J/\Delta n_h=R(\Omega,\Omega)/P(\Omega,\Omega)\sim \Omega^{1/z}$. Then the probability,
$Pr(\alpha)$, that the value of a coupling, $J$, being neighbor to a decimated field is $\Omega<J<\alpha \Omega$ with $0<\alpha<1$
is given by $Pr(\alpha)=1-\alpha^{\tilde{R}\Omega}$, which goes to zero for any non-zero $\alpha$, since
$\tilde{R}\Omega \sim \Omega^{1/z}$ at the fixed point. Consequently the decimations in Eq.(\ref{htilde})
and the related RG equation in Eqs.(\ref{rgJ}) and (\ref{rgh}) are indeed exact.

The relation in Eq.(\ref{zexact}) can be recovered starting from the exact expression for the
surface magnetization, $m_s$, on a chain with a fixed spin at site $L+1$
\cite{peschel,bigpaper}:
\be
{1 \over m_s^2}=1+\sum_{l=1}^L \prod_{i=1}^l \left({h_i \over J_i}\right)^2\;.
\label{peschel}
\ee
This type of expression is the so-called Kesten-variable in the mathematical literature\cite{kesten} and 
the corresponding probability distribution is singular at $m_s=0$ in the thermodynamic limit for
$[\ln J]_{\rm av} > [\ln h]_{\rm av}$\cite{kesten,theo}. Then for the distribution of $\ln m_s$ we have
the following singularity:
\be
P(\ln m_s) \sim m_s^{1/\tilde{z}},\quad m_s \to 0\;,
\label{ms}
\ee
where $\tilde z$ is the solution of Eg.(\ref{zexact}) in terms of the dual variables.
The physical origine of the small $m_s$ tail of the distribution is due to such samples which have a weakly
coupled domain (WCD), which effectively cuts the system inty two very weakly interacting parts and thus
reduces the surface order enormously. In the dual system in the paramagnetic phase the dual object to a
WCD is a strongly coupled domain (SCD) which results in a very small energy gap, $\epsilon$. Thus in the
tails of the distributions, $m_s$ and $\epsilon$ are dual quantities. This remark
indicates that the dynamical exponent found by the RG calculation
in Eq.(\ref{zexact}) is indeed exact\cite{remark}.

Next, we consider general random quantum spin chains with a critical IRFP and analyze the structure of the
RG equations close to the line of semi-critical fixed points, thus as $\Omega \to 0$. As for the RTIM, the
decimation for fields and couplings is asymmetric and for $\Omega \to 0$ exclusively fields are decimated out, which are
typically infinitly stronger, than the neighboring couplings. Therefore the RG decimation equations in Eq.(\ref{htilde})
are asymptotically exact. The second point is to show that the dynamical exponent stays invariant along the
RG trajectory, even though in the starting phase the RG equations are approximative.
For this we consider the low energy tail of the
distribution function of the first gap, $P_1(\ln \epsilon)$, which involves in analogy to Eq.(\ref{ms}) the exponent
$z$, and use the scaling result of
Ref\onlinecite{ijr}. This states that the probability distribution of the second, third, etc. gaps are
related to $P_1(\ln \epsilon)$ as $P_2(\ln \epsilon) \sim P_1^2(\ln \epsilon)$,
$P_3(\ln \epsilon) \sim P_1^3(\ln \epsilon)$, etc, due to the fact that for a small second, third gap one needs two,
three independent SCD-s and the corresponding probabilities are multiplied. In the RG decimation
the SCD-s are only eliminated through coupling decimation, since their couplings are stronger than
the average fields. If at some time a SCD with a small gap, $\epsilon$, is eliminated then in the probability
distribution,
$P_1(\ln \epsilon)$, one should consider the former second gap and use the corresponding conditional probability,
$P_1(\ln \epsilon) \to P_2(\ln \epsilon)/P_1(\ln \epsilon) \sim  P_1(\ln \epsilon)$. Thus the small energy
tail of the gap-distribution
and consequently the dynamical exponent remains invariant under the renormalization procedure. The previously obtained
exact results for the RTIM give strong support for the validity of these phenomenological considerations.

For a numerical demonstration of the validity of the above statement we considered two random quantum spin chains,
the dimerized Heisenberg (XXX) chain and the q-state RQPM, both having a set of RG equations very similar to that of the
RTIM in Eq.(\ref{htilde}). For the dimerized XXX chain $J$ and $h$ in Eq.(\ref{htilde}) are replaced by the
Heisenberg couplings at odd and even positions, $J_o$ and $J_e$, respectively, and the parameter takes the value
$\kappa=2$\cite{mdh,fisherxx}. The distance from the critical point is measured similarly to
Eq.(\ref{delta}). For the q-state RQPM fields and couplings play analogous r\^ole
as for the RTIM, the quantum control-parameter is given in Eq.(\ref{delta}), whereas
$\kappa$ takes the value $\kappa=q/2$\cite{senthil}. We note that the RG equations for the XXX-chain and the
$q=4$ state RQPM are identical.

\begin{figure}    
\hspace{0mm} \epsfxsize= 7,5cm \epsfbox{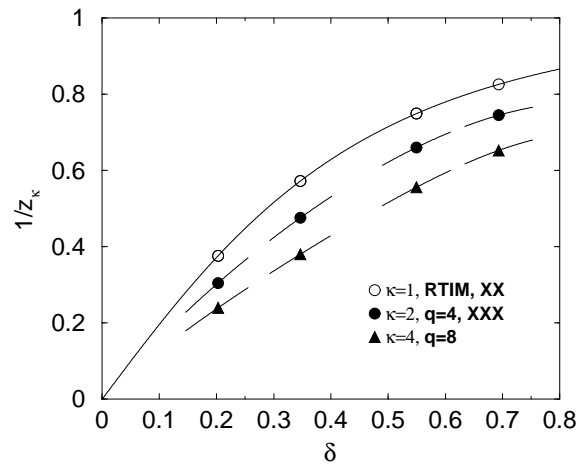}  
\vspace{5mm}

\caption{Dynamical exponents from numerical iteration of the RG-equations in Eq.(\ref{htilde}) for different values of the 
parameter $\kappa$. For $\kappa=1$, which corresponds to the RTIM and the random dimerized XX-chain the exact
result is given by the full line, for $\kappa=2$ and $4$ the broken lines are guide to the eye.}
\label{fig:1}
\end{figure}

At the critical point the RG-equations for $1 \le \kappa < \infty$ has been solved by Senthil and
Majumdar\cite{senthil} with the result that $\kappa$ is an irrelevant variable and the IRFP is the same as
for the RTIM. In the Griffiths-phase we could not find a complete solution of the RG equations, in spite of the
close similarity to that of the RTIM. We could, however, show that up to an accuracy of $O(\Omega^{1/z})$ the solution
is of the form of Eqs.(\ref{solu}) and thus there is infinite randomness along the line of fixed points.
The $z$ exponent, however, does depend on the parameter $\kappa$, since the validity of the condition in
Eq.(\ref{zexact}) is limited to $\kappa=1$, thus in general $z=z_{\kappa}(\delta)$.

We have calculated the dynamical exponent by a numerical implementation
of the RG scheme over 50000 samples of length $L \le 2^{14}$. Starting with the uniform probability distribution:
$R_0(J)=\Theta(1-J) \Theta(J)$, $P_0(h)=\Theta(h_0-h) \Theta(h)/h_0$ (and analogously for the XXX-chain)
we got the estimates shown in Fig.1:
$1/z_{\kappa}$ is a monotonously decreasing function of $\kappa$ and eventuelly it goes to zero in the whole Griffiths phase
in the limit $\kappa \to \infty$.

\begin{figure}
\vspace{-25mm}
\hspace{-20mm} \epsfxsize= 12,8cm \epsfbox{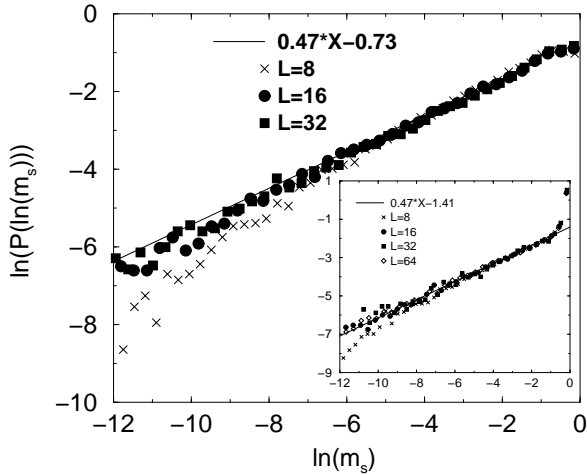}
\vspace{-68mm}
\caption{Probability distribution of the logarithm of the surface magnetization of the dimerized random
XXX-chain with $\delta=-(\ln 2)/2$ for different finite systems calculated by the DMRG method. The full
straigth line with a slope $1/z(\delta)=.47$ represents the asymptotic behavior according to the
RG method, as given in Fig. 1. In the inset the same quantity is shown for the $q=4$ state RQPM
at the same distance from the critical point. The asymptotic slope of this distribution can be well
fitted by the same exponent as for the XXX-model. }
\label{fig:2}
\vspace{0mm}
\end{figure}

The dynamical exponents of the XXX-chain and the RQPM are also calculated directly from the asymptotic behavior of the
distribution of the surface magnetization, as given in Eq.(\ref{ms}). For the numerical calculations of the
surface magnetizations we used the DMRG method for rather large finite chains with $L \le 64$ and considered
some 20000 samples. We found an overall agreement between the dynamical exponents calculated by the two methods.
As a demonstration we show in Fig. 2 the distribution of $m_s$ for the XXX-chain, compared with
that of the $q=4$ state RQPM, where for both models we are at the same distance from the transition point.
As seen in Fig. 2 the asymtotic behavior of the two distributions is identical, as expected on the RG basis, since
$\kappa=2$ for both models. Furthermore the dynamical exponents agree very well with those calculated by the RG method.

To conclude, we have shown in this paper that for random quantum spin chains the RG method of Ma, Dasgupta and Hu
is asymptotically exact for large times, i.e. along the line of semi-critical fixed points. Consequently the dynamical
exponent in the Griffiths phase calculated by the RG-method is exact. This result will hopefully stimulate new
efforts to solve the RG equations for different systems analytically in 1d 
also outside the critical point, and in
higher dimensions to perform numerical calculations and get more precise estimates for the Griffiths-McCoy
singularities.

Acknowledgment: This work has been
supported by the Hungarian National Research Fund under grant No OTKA
TO23642, TO25139, F/7/026004, MO28418  and by the Ministery of Education under grant No. FKFP 0596/1999. We are indebted to
E. Carlon for his participation in the early stages of the DMRG calculations. F. I. is grateful
to H. Rieger and L. Turban for previous cooperation in the project and for useful discussions.

\end{multicols}
\end{document}